\documentclass[prb,twocolumn,showpacs,preprintnumbers,amsmath,amssymb,amsfonts]{revtex4}

\usepackage{graphicx}
\usepackage[dvips]{color}
\usepackage{dcolumn}
\usepackage{bm}
\usepackage{setspace}
\usepackage{subfigure}

\begin{document}

\title{Circularly polarized soft x-ray diffraction study of helical magnetism in hexaferrite}

\author{AM Mulders$^{1,2}$, SM Lawrence$^{1}$, AJ Princep$^{1}$, U Staub$^{3}$, Y Bodenthin$^{3}$, M Garcia-Fernandez$^{3}$, M Garganourakis$^{3}$, J Hester$^{2}$, R Macquart$^{4}$ and CD Ling$^{2,4}$}

\affiliation{$^{1}$ Department of Imaging and Applied Physics, Curtin University of Technology, Perth, WA 6845,  Australia}
\affiliation{$^{2}$ The Bragg Institute, Australian Nuclear Science and Technology Organisation, Lucas Heights, NSW 2234, Australia} 
\affiliation{$^{3}$ Swiss Light Source, Paul Scherrer Institut, 5232 Villigen PSI, Switzerland}
\affiliation{$^{4}$ School of Chemistry, The University of Sydney, Sydney, New South Wales 2006, Australia}

\begin{abstract}
Magnetic spiral structures can exhibit ferroelectric moments as recently demonstrated in various multiferroic materials. In such cases the helicity of the magnetic spiral is directly correlated with the direction of the ferroelectric moment and measurement of the helicity of magnetic structures is of current interest. Soft x-ray resonant diffraction is particularly advantageous because it combines element selectivity with a large magnetic cross section. 
We calculate the polarization dependence of the resonant magnetic x-ray cross section (electric dipole transition) for the basal plane magnetic spiral 
in hexaferrite Ba$_{0.8}$Sr$_{1.2}$Zn$_2$Fe$_{12}$O$_{22}$ and deduce its domain population using circular polarized incident radiation.
We demonstrate there is a direct correlation between the diffracted radiation and the helicity of the magnetic spiral.

\end{abstract}

\pacs{75.25.-j, 61.05.cc, 75.50.Dd, 75.84.+t}

\maketitle

Magnetic spiral structures can exhibit ferroelectric moments as recently demonstrated in various
multiferroic materials \cite{1,2,3,kimura_prl_2005}. In such cases the helicity of the magnetic spiral is directly correlated with the direction of the ferroelectric moment and measurement of the helicity of magnetic structures is of
current interest.
Magnetic spiral structures have been observed directly with neutron diffraction (ND) and resonant x-ray diffraction (RXD) \cite{hannon_prl_1988} as their super structure gives rise to satellite reflections. With polarized neutron diffraction the chirality of the magnetic structure can be determined, as was first predicted by Blume \cite{blume_pr_1963} and achieved by Siratori \cite{siratori_jpsj_1980}. Recently this has been particularly insightful for the study of ferroelectric magnetic spiral structures in TbMnO$_3$ \cite{yamasaki_prl_2007}, LiCu$_2$O$_2$ \cite{seki_prl_2008} and CuFe$_{1-x}$Al$_x$O$_2$ \cite{nakajima_prb_2008}
observing that the chirality of the magnetic structure is manipulated with applied electric field.
With circular polarized non-resonant x-ray diffraction the chiral magnetic domain population in holmium has been determined \cite{sutter_prb_1997} and, very recently, polarization analysis has been used to study the cycloidal magnetic domains in multiferroic TbMnO$_3$ in its ferroelectric phase \cite{fabrizi_prl_2009}.

An advantage of RXD is that via tuning the incident energy to a particular absorption edge, element specific magnetism is observed. In the case of transition metals, the $L_{2,3}$ edge is particularly insightful because the core electron is excited from the core 2p to the 3d valence states and the (empty) magnetic states are directly probed. The magnetic cross section is significant compared to the charge cross section and soft x-ray resonant diffraction has emerged as a very valuable technique with which to study magnetic and orbital order in transition metal oxides, in particular to distinguish between charge, orbital and magnetic order \cite{staub_prb_2005, wilkins_prb_2005, herrero_prb_2006, scagnoli_prb_2006a}. 
Correlation between the RXD intensity and the helicity of the magnetic spiral has been demonstrated by imaging of the spiral domains in holmium \cite{lang_jap_2004} but a quantitative analysis of the diffracted intensities is missing. 

In this paper we calculate the polarization dependence of the RXD cross section and deduce the domain population of the magnetic spiral structure in hexaferrite using variably polarized incident radiation at the Fe $L_3$-edge ($\lambda$=17.45 \AA).
We demonstrate that XRD is well suited to study magnetoelectric coupling in multiferroic materials that exhibit magnetic spiral components.

\begin{figure}[b]
\vspace{-.5cm}
\includegraphics[width=14pc,angle=90]{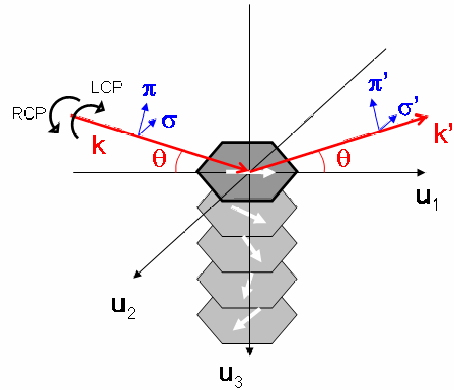}
\caption{\label{fig_coor} Experimental set-up with coordinate system used for calculating the polarization dependent resonant cross section. The basal plane magnetic spiral with helicity $S^+$ is indicated.}
\end{figure}

\begin{figure}[b]
\vspace{-1.6cm}
\includegraphics[width=14.5pc,angle=-90]{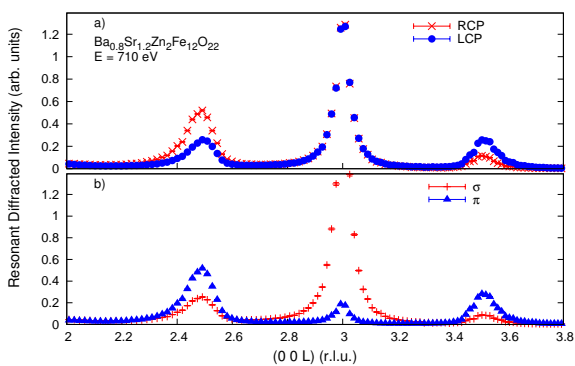}
\caption{\label{fig_all} Resonant diffraction of hexaferrite at T = 35~K recorded with circular (a) and linear (b) polarized incident radiation at the Fe $L_3$ edge. Note the asymmetry in intensity of the two magnetic satellites neighboring the (003) reflection recorded with RCP (red crosses) and LCP (blue filled spheres). $\sigma$ (blue filled triangles) and $\pi$ (red plusses) diffraction does not distinguish between positive and negative helicity of the magnetic spiral. }
\end{figure}

\begin{figure}[b]
\vspace{-1.3cm}
\includegraphics[width=22pc,angle=-90]{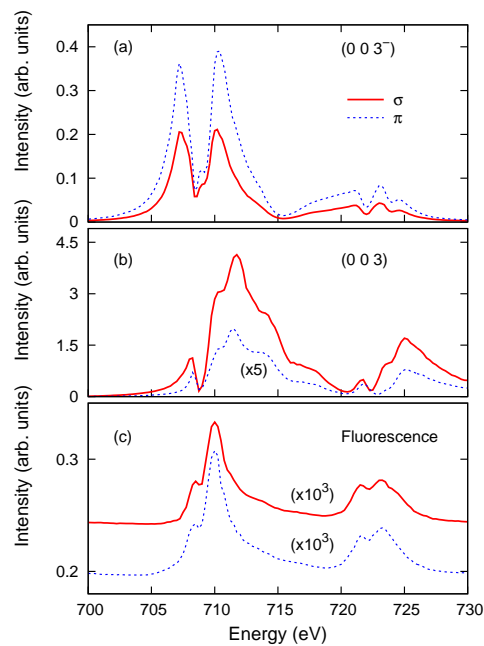}
\caption{\label{fig_e} Energy dependence of  the (003$^-$) magnetic satellite (a) compared to  the (003) structural reflection (b) and the fluorescence yield (c), recorded with $\sigma$ (solid red) and $\pi$ (dotted blue) linear polarized incident radiation. Note the scaling factors used. The spectra were recorded at T $\sim$ 90~K, 132~K and 138~K, respectively and are not corrected for absorption.
}
\end{figure}

The helical magnetic structure of Ba$_{0.5}$Sr$_{1.5}$Zn$_2$Fe$_{12}$O$_{22}$ hexaferrite \cite{enz_jap_1961} has been studied with neutron scattering \cite{momozawa_jpsj_1986,momozawa_jpsj_1993, utsumi_jpsj_2007} and polarized x-ray diffraction \cite{tsuji_jpsj_1996} and is characterized by (0 0 l$^\pm$) satellites with l$^\pm$ = 3n$\pm 3\tau$. 
Competition among superexchange interactions leads to a distorted helimagnetic structure consisting of large and small ferrimagnetic bunches with moments aligned in the $ab$ plane and modulation wave vector (0 0 $\tau$). Electric polarization arises under applied magnetic field in the intermediate-III magnetic
phase, characterized by $\tau$ = 0.5.

Hexaferrite single crystals were grown in a manner similar to that given by Momozawa et al.  \cite{58}. Reactants were mixed in the molar ratio 4.92\% BaCO$_3$, 14.77\% SrCO$_3$, 19.69\% ZnO, 53.61\% Fe$_2$O$_3$, 7.01\% Na$_2$CO$_3$ flux and heated in a Pt crucible to 1420 $^\circ$C at a rate of 20 $^\circ$C/min and left there for 20 hours.  Rapid temperature cycling  \cite{58} was employed to remove impurity crystals followed by slow cooling (0.2 $^\circ$C/h) between 1185 $^\circ$C and 1155 $^\circ$C to improve crystal size. 

Elemental analysis with inductively-coupled plasma atomic emission spectroscopy gave a stoichiometry of Ba$_{0.8}$Sr$_{1.2}$Zn$_2$Fe$_{12}$O$_{22}$, with small variations in Ba/Sr ratio depending on the single crystal.
Magnetization measurements were performed with a Quantum Design 7T MPMS at the magnetics laboratory at the University of Western Australia. The single crystals exhibited consecutive magnetization steps as function of applied magnetic field in the $ab$ plane, similar to results reported earlier \cite{momozawa_jpsj_1993,kimura_prl_2005}. 
Single crystal neutron diffraction was performed at the Wombat diffractometer at OPAL with $\lambda$ = 2.955 $\AA$ and showed temperature dependent incommensurate $\tau$ for $B$ = 0 T, $\tau$ = 0.5 for 0.5 $\leq$ $B$ $\leq$ 2.0 T and $\tau$ = 1 for $B$ $\leq$ 2.25 T (at T = 100 K).
Soft x-ray resonant diffraction was performed at the RESOXS end-station of the SIM beamline at the Swiss Light Source of the Paul Scherrer Institut \cite{urs_jsr_2008}.
The elliptical twin undulator UE56 (Apple II) of the beamline can produce variable linear and circular polarization. 
The hexaferrite sample was mounted with the $c$-axis in the scattering plane and the $a$-axis perpendicular to the scattering plane (see Fig. \ref{fig_coor}) and its temperature was regulated between 15 K and 325 K.

X-ray diffraction from magnetic spiral structures gives rise to satellite reflections which have polarization dependent magnetic scattering cross sections \cite{blume_prb_1988}. The intensity from magnetic moments is proportional to $(\hbar\omega/mc^2)^2$ and is generally weak. However, this is enhanced by orders of magnitude when the x-ray energy is tuned to an absorption edge that excites a core electron to the empty valence states.
In case of transition metals this is the $L_3$ edge (electric dipole transition) and the resonant magnetic scattering amplitude of the magnetic satellites equals \cite{hannon_prl_1988}
\begin{equation}
f_{\varepsilon'\varepsilon}^{XRES} = -\frac{3}{4 k} i ({\bm \hat{\varepsilon}'} \times {\bm \hat{\varepsilon}}) \cdot {\bf \hat{z}}_j [F_{11}-F_{1-1}]
\label{eq_f}
\end{equation}
where ${\bm \varepsilon}$ and ${\bm \varepsilon'}$ are the polarization of the incident and diffracted radiation, ${\bf \hat{z}}_j$ the quantization axis of the magnetic moment at atom $j$ and $F_{11}, F_{1-1}$ are the atomic properties of the initial and excited state of the Fe ion which is related to the $3d$ magnetic moment and the overlap integral. 
${\bf q = k - k'}$ is the wavevector transfer and ${\bf k}$ and ${\bf k'}$ are the wavevector of the incident and diffracted radiation, respectively. The sum over magnetic ions $j$ gives the magnetic structure factor $M_{\varepsilon'\varepsilon}^{XRES} = \sum_j \exp(i {\bf q} \cdot {\bf r}_j) f_{\varepsilon'\varepsilon}^{XRES}$.

Hill and McMorrow have formulated the resonant x-ray cross section in terms of linear polarized radiation \cite{hill_aca_1996}. 
Recently, Lovesey {\it et al} formulated the resonant diffracted intensity for circular polarized radiation in terms of the scattering factors for linear polarization \cite{lovesey_jpcm_2008}. 
Without polarization analysis the diffracted intensities $I_{\varepsilon}^{XRES}$ equal
\begin{eqnarray}
I_{\sigma}^{XRES} &=& |M_{\sigma' \sigma}|^2 +  |M_{\pi'\sigma}|^2\\
I_{\pi}^{XRES} &=& |M_{\sigma'\pi}|^2 + |M_{\pi'\pi}|^2\\
I_{\chi}^{XRES} &=& \frac{1}{2} \left( |M_{\sigma' \sigma}|^2 +  |M_{\pi'\sigma}|^2 + |M_{\sigma'\pi}|^2 + |M_{\pi'\pi}|^2 \right) \nonumber \\ && \ \ \ + \chi {\rm Im}\{ M_{\sigma' \sigma}M_{\sigma'\pi}^* + M_{\pi'\sigma}M_{\pi'\pi}^* \} \label{eq_circ}
\label{eq_lovesey}
\end{eqnarray}
where $\pi$ polarization is parallel to the scattering plane and $\sigma$ polarization is perpendicular to the scattering plane (see Fig. \ref{fig_coor}). $\chi = +1$ indicates right circular polarized (RCP) and $\chi = -1$ left circular polarized (LCP) of the incident beam. The last term of eq. (\ref{eq_circ}) generally vanishes but can be non zero as demonstrated for the case of the enantiomorphic screw axis in quartz \cite{tanaka_prl_2008}.

\begin{figure}[b]
\vspace{-1.8cm}
\includegraphics[width=13pc,angle=-90]{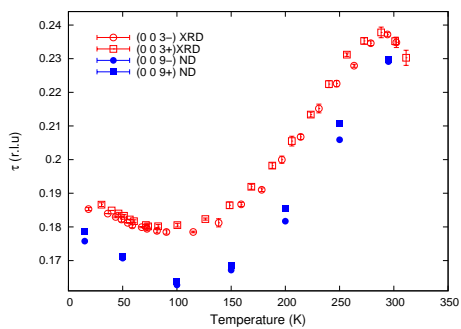}
\vspace{-0.5cm}
\caption{\label{fig_temp} Wavevector of the magnetic spiral in hexaferrite as a function of temperature deduced from the resonant magnetic satellite diffraction (open symbols) and neutron diffraction (solid symbols).
The difference in the magnitude of $\tau$ for XRD and ND is attributed to stoichiometry variation between the single crystals used in the experiments.}
\end{figure}

We approximate the magnetic structure of hexaferrite with a single magnetic moment that rotates either clockwise (positive helicity $S^+$) or anticlockwise (negative helicity $S^-$) for consecutive atoms along the $c$-axis. For a basal plane magnetic spiral with modulation wavevector (0 0 ${\bm \tau}$) the moment direction for $S^+$ and $S^-$ equals ${\bf \hat{z}}_j^\pm $=$ {\bf \hat{u}}_1 \cos ({\bm \tau} \cdot {\bf r}_j) \pm {\bf \hat{u}}_2 \sin ({\bm \tau} \cdot {\bf r}_j) $=$ \frac{1}{2} [ {\bf \hat{u}}_\mp \exp(i {\bm \tau} \cdot {\bf r}_j) + {\bf \hat{u}}_\pm \exp(-i {\bm \tau} \cdot {\bf r}_j)]$ with ${\bf \hat{u}}_\pm = {\bf \hat{u}}_1 \pm i{\bf \hat{u}}_2$. The unit vectors ${\bf \hat{u}}_1$, ${\bf \hat{u}}_2$ and ${\bf \hat{u}}_3$ define the coordinate system with respect to the diffraction plane. ${\bf \hat{u}}_1$ is parallel to ${\bf k'}+{\bf k}$, ${\bf \hat{u}}_2$ is perpendicular to the scattering plane, and ${\bf \hat{u}}_3$ is  parallel to ${\bf k'}-{\bf k}$ (see Fig. \ref{fig_coor}). In this experiment the c-axis of hexaferrite is aligned along ${\bf \hat{u}}_3$. The cross terms of polarization ${\bm \hat{\varepsilon}}' \times {\bm \hat{\varepsilon}}$ in eq.( \ref{eq_f}) equal ${\bm \sigma}'\times{\bm \sigma}$ = 0, ${\bm \sigma}'\times{\bm \pi} = {\bf \hat{k}}$, ${\bm \pi}'\times{\bm \sigma} = -{\bf \hat{k}}'$ and ${\bm \pi}'\times{\bm \pi} = {\bf \hat{k}}' \times {\bf \hat{k}}$. Consequently, the polarization dependent magnetic structure factors are
\begin{eqnarray}
M_{\sigma' \sigma}&=&0\\
M_{\pi'\sigma}&=&-\frac{3}{4 k} i\sum_j (-{\bf \hat{k}'} \cdot {\bf \hat{z}}_j^\pm)[F_{11}-F_{1-1}]\exp(i {\bf q} \cdot {\bf r}_j)\\
M_{\sigma'\pi}&=&-\frac{3}{4 k} i\sum_j ({\bf \hat{k}} \cdot {\bf \hat{z}}_j^\pm)[F_{11}-F_{1-1}]\exp(i {\bf q} \cdot {\bf r}_j)\\
M_{\pi'\pi}&=&-\frac{3}{4 k}i\sum_j({\bf \hat{k}'}\times{\bf \hat{k}})\cdot{\bf \hat{z}}_j^\pm[F_{11}-F_{1-1}] \exp(i {\bf q} \cdot {\bf r}_j) \nonumber \\
\end{eqnarray}

\noindent Using ${\bf \hat{k}} = {\bf \hat{u}}_1 \cos \theta + {\bf \hat{u}}_3 \sin \theta$, ${\bf \hat{k}'} = {\bf \hat{u}}_1 \cos \theta - {\bf \hat{u}}_3 \sin \theta$ and ${\bf \hat{k}'} \times {\bf \hat{k}} = - {\bf \hat{u}}_2 \sin 2\theta$ gives for the resonant intensity of the magnetic satellites
\begin{eqnarray}
I_{\sigma, S^\pm}^{XRES}&\propto&\cos^2\theta [F_{11}-F_{1-1}] \delta({\bf q} \pm {\bm \tau}) \label{eq_sigma}\\
I_{\pi, S^\pm}^{XRES}&\propto&(\cos^2\theta + \sin^2 2\theta) [F_{11}-F_{1-1}] \delta({\bf q} \pm {\bm \tau}) \label{eq_pi}\\
I_{\chi,S^+}^{XRES}&\propto&(\cos^2\theta + \frac{1}{2}\sin^22\theta \nonumber \\ 
&&\mp \chi  \cos \theta \sin 2 \theta) [F_{11}-F_{1-1}]^2 \delta({\bf q} \pm {\bm \tau}) \label{eq_splus}\\
I_{\chi,S^-}^{XRES}&\propto&(\cos^2\theta + \frac{1}{2}\sin^22\theta \nonumber \\
&&\pm \chi  \cos \theta \sin 2 \theta) [F_{11}-F_{1-1}]^2 \delta({\bf q} \pm {\bm \tau}) \label{eq_sminus}
\end{eqnarray}
The diffracted intensity for linear polarization is independent of the helicity of the magnetic spiral, 
in contrast to the diffracted intensity for circular polarization which is distinct for S$^+$ and S$^-$. The term proportional to $ \cos \theta \sin 2 \theta$ is opposite in sign for the satellites at (0 0 l$^-$) and (0 0 l$^+$) and, for each satellite, the difference in resonant intensity for RCP and LCP incident radiation is proportional to $2 \cos \theta \sin 2 \theta [F_{11}-F_{1-1}]^2$.
This demonstrates a direct correlation between diffracted intensity of circularly polarized x-rays and the helicity of the magnetic spiral.
The deduced intensities (eqs. \ref{eq_sigma}-\ref{eq_sminus}) are independent of rotation of the sample around the scattering factor ${\bf q}$.

\begin{table}[t]
\vspace{-.3cm}
\caption{\label{table}Experimental (exp.) and calculated (calc.) intensities for the magnetic satellites observed in hexaferrite using 77 \% of magnetic domains with $S^+$ and 23 \% of magnetic domains with $S^-$. A single scaling factor was used between experimental and calculated intensities.}
\begin{tabular}{@{}lllll}
\hline 
incident&\multicolumn{2}{c}{(003$^-$)}&\multicolumn{2}{c}{(003$^+$)}\\
polarization~~~~& exp.~~~~ & calc.~~~~ & exp.~~~~ & calc.~~~~\\
\hline
RCP&0.50&0.52&0.11&0.10\\
LCP&0.24&0.25&0.25&0.23\\
$\sigma$&0.23&0.26&0.08&0.08\\
$\pi$&0.50&0.51&0.27&0.25\\
\hline 
\end{tabular}
\end{table}

Fig. \ref{fig_all} illustrates that (0 0 3$^+$) is most intense for LCP and (0 0 3$^-$) is most intense for RCP radiation. In contrast, linear polarization of the incident radiation does not result in such asymmetry. The diffracted intensity for $\pi$ radiation is stronger than for $\sigma$ radiation consistent with eqs. (\ref{eq_pi}) and (\ref{eq_sigma}). 
The resonant Bragg intensity as a function of incident energy for the (003$^\pm$) satellite is distinct from that of the (003) reflection and orders of magnitude stronger than the fluorescence yield (Fig. \ref{fig_e}).
Table \ref{table} compares the observed and calculated intensities demonstrating good agreement,
77\% exhibits a magnetic spiral with $S^-$ and 23 \% a magnetic spiral with $S^+$ within the illuminated sample area of $\sim$1 mm$^2$. 

Fig \ref{fig_temp} shows the gradual modulation of the wavevector of the magnetic spiral with increasing temperature.  
Variations in stoichiometry result in different $\tau$ due to modification of the superexchange interaction.
The general trend is similar to that observed for Ba$_{0.5}$Sr$_{1.5}$Zn$_2$Fe$_{12}$O$_{22}$ \cite{momozawa_jpsj_1986,utsumi_jpsj_2007}.  

RXD from the magnetic spiral is intense and on average about a factor of two smaller than the diffracted intensity at (003). This is much stronger than the non resonant x-ray diffracted intensity reported for Ba$_{0.5}$Sr$_{1.5}$Zn$_2$Fe$_{12}$O$_{22}$ \cite{tsuji_jpsj_1996}. 
The long wavelength makes soft x-ray resonant diffraction particularly suited to study magnetic structures with large periodicity. Recently, soft x-ray diffraction studies at the Mn $L_3$ edge have measured the magnetic spin structure and order parameter in multiferroic TbMn$_2$O$_5$ demonstrating the magnetoelectric effect arises from non collinear spin moments \cite{okamoto_prl_2007}. An in situ applied electric field demonstrated significant manipulation and the excitation of commensurate magnetic order in multiferroic ErMn$_2$O$_5$ \cite{bodenthin_prl_2008}. Very recently circularly polarized radiation at the Dy $M_5$ resonance has been used to write and read chiral domains in multiferroic DyMnO$_3$ \cite{schierle_arXiv_2009}.

In summary, we present polarization dependent resonant diffraction cross sections for $\sigma$, $\pi$, LCP and RCP incident radiation and determine the spiral magnetic domain population. This demonstrates the potential of using circular polarized soft x-ray resonant diffraction for investigation of long wavelength magnetic structures.
Compared to polarized neutron studies, this method records element specific magnetic helicity and allows to differentiate between distinct magnetic ions within one material. 
The large diffracted intensity, limited sample size requirements, and speed of collection times are an additional advantage.

We acknowledge discussion with Hoyoung Jang. This work was partly performed at the SLS of the Paul Scherrer Institut, Villigen, Switzerland. We thank the beamline staff of X11MA and Robert Woodward from UWA for their support. We acknowledge financial support from the Swiss National Science Foundation, AINSE, Access to Major Research Facilities Programme and the ARC - Discovery Projects (DP0666465).

\end{document}